\documentclass[twocolumn,pra,showpacs,superscriptaddress]{revtex4-1}
\usepackage{amssymb}
\usepackage{amsmath}
\usepackage{graphicx}
\usepackage{epsfig}

\begin{document}

\title{Wave emission and absorption at spectral singularities}
\author{P. Wang}
\affiliation{School of Physics, Nankai University, Tianjin 300071, China}
\author{L. Jin}
\email{jinliang@nankai.edu.cn}
\affiliation{School of Physics, Nankai University, Tianjin 300071, China}
\author{G. Zhang}
\affiliation{School of Physics, Nankai University, Tianjin 300071, China}
\affiliation{College of Physics and Materials Science, Tianjin Normal University, Tianjin
300387, China}
\author{Z. Song}
\affiliation{School of Physics, Nankai University, Tianjin 300071, China}

\begin{abstract}
We studied the critical dynamics of spectral singularities. The system
investigated is a coupled resonator array with a side-coupled loss (gain)
resonator. For a gain resonator, the system acts as a wave emitter at
spectral singularities. The reflection probability increased linearly over
time. The rate of increase is proportional to the width of the incident wave
packet, which served as the spectral singularity observer in the experiment.
For a lossy resonator, the system acts as a wave absorber. The emission and
absorption states at spectral singularities coalesce in a finite parity-time
($\mathcal{PT}$) symmetric system that combined by the gain and loss
structures cut from corrresponding scattering systems at spectral
singularities; in this case, the $\mathcal{PT}$-symmetric system is at an
exceptional point with a $2\times 2$ Jordan block. The dynamics of the $%
\mathcal{PT}$-symmetric system exhibit the characteristic of exceptional
points and spectral singularities.
\end{abstract}

\maketitle


\section{Introduction}

\label{introduction}

A non-Hermitian Hamiltonian can possess peculiar features that have no
counterpart in a closed Hermitian system. For example, nonreciprocal
dynamics, have been observed in experiments~\cite{Observe}. Research
indicates that the combination of a magnetic field and non-Hermitian
potential has an unexpected effect on particle transport behavior~\cite%
{Longhi2015OL,LXQ}. The discovery of non-Hermitian Hamiltonians with
parity-time ($\mathcal{PT}$) symmetry and a real spectrum~\cite{Bender} has
furthered research on the complex extension of quantum mechanics on a
fundamental level~\cite%
{Ann,JMP1,JPA1,JPA2,PRL1,JMP2,JMP3,JMP4,JPA3,JPA4,Ali1,JPA5,LJin09,LJin10}.
In recent years, following revelations of their possible physical relevance
by the pioneering work of Ali Mostafazadeh, the spectral singularities of
non-Hermitian systems have received considerable attention~\cite%
{PRA1,PRB1,Ali3,PRA3,JMP5,PRD1,PRA4,PRA5,PRA6}. Most studies have focused on
non-Hermitian systems with $\mathcal{PT}$ symmetry~\cite%
{PRA2,JPA6,Ali3,PRA13,prd2,prd3,prd4,prd5,prd6,prd7,prd8} and non-Hermitian
hopping amplitude~\cite{PRA14,ZXZ,S. Longhi14,LGR,Longhi2015SR,Longhi2015PRB}%
.

Spectral singularities are divergences in the continuous spectrum of
scattering systems~\cite{PRA2}, and differ from exceptional points~\cite%
{HeissEP,RotterEPJPA,RotterEPPRE}. Spectral singularities are attributed to
non-Hermitian terms, leading to transmission and (or) reflection
coefficients that tend to infinity for an incidence. The steady-states at
spectral singularities in scattering systems have the form of propagating
plane wave emission and (or) absorption~\cite{PRA14}. A spectral singularity
can represent lasing with zero linewidth~\cite{PRL3,AhmedSS}. By using Fano
resonance in a $\mathcal{PT}$-symmetric system with a pair of side-coupled
balanced gain and loss resonators, spectral singularity induces
unidirectional lasing is determined~\cite{XZhangUSS}, where spectral
singularities exhibit nonreciprocity. Complex potential induced spectral
singularities in scattering system are intensively investigated~\cite%
{PRA3,Ali3,AliJPA,AliSS2011JPA,PRA1,AhmedSS,XZhangUSS,HeissSS}.

Wave packet dynamics at the spectral singularities in a
Friedrichs-Fano-Anderson model was predicted~\cite{PRB1}. In this paper, we
qualitatively studied the dynamic phenomena of wave emission and absorption
in the coupled resonator array through a side-coupled gain (loss) resonator
using an exact solution of a concrete tight-binding system. Spectral
singularities occur when the wave vector matches the coupling strength and
gain (loss) rate. The critical behavior associated with the physics of the
spectral singularity was investigated, including the wave emission and the
perfect absorption of the incident wave. The perfect absorption corresponded
to a coherent perfect absorption induced by loss; for wave emission, the
reflection probability at the gain resonator increases linearly with time.
The secular wave emission forms a platform with wave amplitude being
proportional to the full width at half maximum (FWHM) of the incident wave
packet. This can serve as the spectral singularity witness in the
experiment. We combined the leads of two semi-infinite scattering systems
into a $\mathcal{PT}$-symmetric system. We demonstrated that the $\mathcal{PT%
}$-symmetric system was at an exceptional point with a $2\times 2$ Jordan
block when the two semi-infinite scattering systems were at spectral
singularities. Thus, we linked the spectral singularity to the exceptional
point. We calculated the dynamics of a Gaussian wave in the $\mathcal{PT}$%
-symmetric system. A quadratic probability increase was seen for the long
time scale. For the short time interval, the probability revealed the
emission and absorption behavior of the spectral singularities.

This study is organized as follows. In Sec.~\ref{Model}, we present the
model setup and the solutions. In Sec.~\ref{Spectral singularity}, the
spectral singularity of the Hamiltonian is discussed. In Secs.~\ref{Emission}
and~\ref{Absorption}, we qualitatively analyze the wave emission and perfect
absorption of a Gaussian wave packet at the spectral singularities. The
scaling law of a wave emission is investigated in detail. In Sec.~\ref{PT},
we compose a $\mathcal{PT}$-symmetric system at the exceptional point using
two semi-infinite scattering systems at the spectral singularities. We
demonstrate that the spectral singularities absorption and emission states
coalesce in the $\mathcal{PT}$-symmetric system. Quadratic increases of the
wave probability in the $\mathcal{PT}$-symmetric system are shown to reflect
that the system is at the exceptional point, consisting of a $2\times 2$
Jordan block. Finally, we present a summary in Sec.~\ref{Summary}.

\section{Model}

\label{Model}

\begin{figure}[tbp]
\includegraphics[bb=45 230 565 727, width=7 cm, clip]{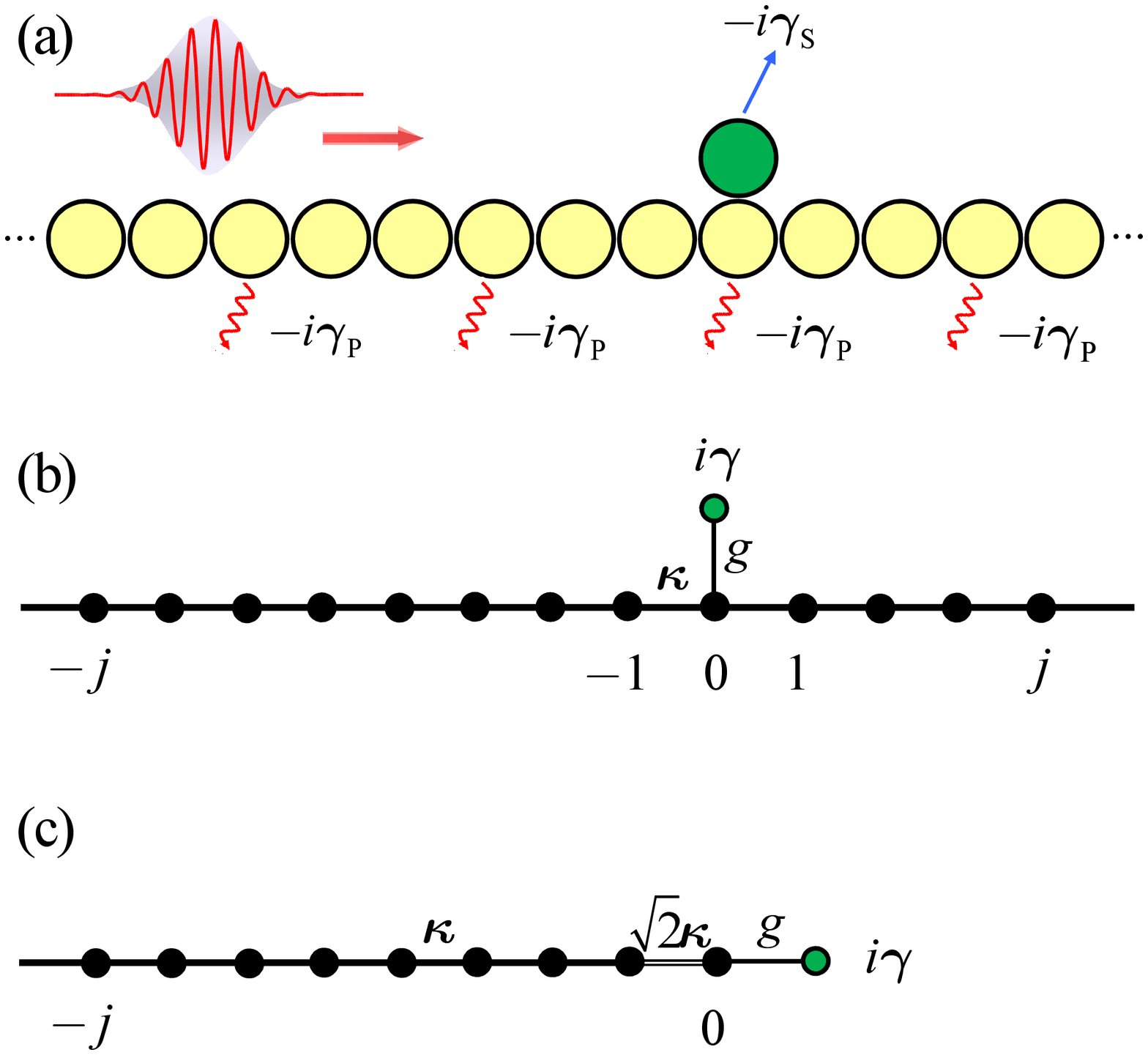}
\caption{(Color online) (a) Schematic configuration for the coupled
resonator system. An array of coupled single-mode resonators with a
resonator side-coupled at the center. Initially, a photon wave packet moves
from the left to the center. Each resonator on the chain array has a decay
rate $\protect\gamma _{\mathrm{P}}$ and the side-coupled resonator has a
decay rate $\protect\gamma_\mathrm{S}$. (b) Schematic illustration of an
equivalent system with a discrete tight-binding model. A tight-binding chain
with uniform coupling, side-coupled to a resonator with an imaginary on-site
potential $i\protect\gamma $. (c) After symmetric and anti-symmetric
combination of the uniform chain, we obtained a semi-infinite chain array
with gain (loss) resonator coupled at the end.} \label{fig1}
\end{figure}

A one-dimensional coupled resonator array, consisting of a side-coupled
resonator is schematically illustrated in Fig.~\ref{fig1}(a).\textbf{\ }The
dynamics of the system are characterized by the coupled mode theory~\cite%
{XZhangUSS}, and the system is equivalently described by the Hamiltonian
\begin{eqnarray}
H &=&H_{0}-ga_{0}^{\dag }a_{\mathrm{S}}-ga_{\mathrm{S}}^{\dag }a_{0}+i\gamma
a_{\mathrm{S}}^{\dag }a_{\mathrm{S}},  \label{H} \\
H_{0} &=&-\kappa \sum_{l=-\infty }^{+\infty }(a_{l}^{\dag
}a_{l+1}+a_{l+1}^{\dag }a_{l})-i\gamma _{\mathrm{P}}\sum_{l=-\infty
}^{+\infty }a_{l}^{\dag }a_{l},  \notag
\end{eqnarray}%
where $a_{l}$ ($a_{l}^{\dag }$) annihilates (creates) a photon at the $l$-th
resonator. $\gamma _{\mathrm{P}}$ and $\gamma _{\mathrm{S}}$\ are the
dissipation rates of the cavities in the chain array and the side-coupled
resonator, respectively. The relative decay rate is set at $\gamma =\gamma _{%
\mathrm{P}}-\gamma _{\mathrm{S}}$. The chain array is denoted by $H_{0}$.
The term $-i\gamma _{\mathrm{P}}$\ has an effect of overall decay on the
eigenstates of $H_{0}$. On the condition that the overall decay rate is
substantially smaller than the photon tunneling rate, i.e., $\gamma _{%
\mathrm{P}}\ll \kappa $, the overall decay of the amplitudes is negligible
in a finite time scale $t\sim \gamma _{\mathrm{P}}^{-1}$. The side coupling
strength is $g$ and the photon hopping strength for the tunneling between
the adjacent cavities is $\kappa $. The Hamiltonian $H$ is schematically
illustrated in Fig.~\ref{fig1}(b). A candidate realization is using the
on-chip coupled microresonator array produced in optical systems~\cite%
{Vahala1,Vahala2}. In previous works, the net gain was created in the
microresonator produced on silicon wafers by pumping the doped erbium ions,
the additional loss was induced by a chromium-coated tip~\cite{PengNP,
PengScience}.

On the bases of the wave function ansatz and the mirror symmetry of the
Hamiltonian $H$, the solution of the Schr\"{o}dinger equation
\begin{equation}
H\left\vert \psi _{k}^{\pm }\right\rangle =\varepsilon _{k}\left\vert \psi
_{k}^{\pm }\right\rangle ,  \label{S-eq}
\end{equation}%
can be obtained as%
\begin{equation}
\left\langle j\right. \left\vert \psi _{k}^{+}\right\rangle =\left\{
\begin{array}{cc}
\eta _{k}e^{ikj}-\eta _{-k}e^{-ikj}, & \left( j\leqslant -1\right) \\
-\eta _{-k}e^{ikj}+\eta _{k}e^{-ikj}, & \left( j\geqslant 1\right)%
\end{array}%
\right. ,
\end{equation}%
and%
\begin{equation}
\left\langle j\right. \left\vert \psi _{k}^{-}\right\rangle =%
\begin{array}{cc}
\sin \left( kj\right) , & \left( \left\vert j\right\vert \geqslant 1\right)%
\end{array}%
,
\end{equation}%
where $\pm $\ represents the parity of the solution, and%
\begin{equation}
\eta _{k}=2i\kappa \sin k\left( i\gamma +2\kappa \cos k\right) +g^{2}.
\label{eta_k}
\end{equation}%
The spectrum is $\varepsilon _{k}=-2\kappa \cos k,$ which is always real for
a scattering state.

According to the theory of pseudo-Hermitian quantum mechanics~\cite{JPA4}, a
complete biorthonormal set requires the construction of the eigenfunctions
of $H^{\dag }$. Similarly, the solution of the Schr\"{o}dinger equation%
\begin{equation}
H^{\dag }\left\vert \overline{\psi }_{k}\right\rangle =\varepsilon
_{k}\left\vert \overline{\psi }_{k}\right\rangle .  \label{S-eq*}
\end{equation}%
can be obtained by taking $\gamma \rightarrow -\gamma $ from $\left\langle
j\right. \left\vert \psi _{k}^{\pm }\right\rangle $, i.e.,%
\begin{equation}
\left\langle j\right. \left\vert \overline{\psi }_{k}^{+}\right\rangle
=\left\{
\begin{array}{cc}
\eta _{-k}^{\ast }e^{ikj}-\eta _{k}^{\ast }e^{-ikj}, & \left( j\leqslant
-1\right) \\
-\eta _{k}^{\ast }e^{ikj}+\eta _{-k}^{\ast }e^{-ikj}, & \left( j\geqslant
1\right)%
\end{array}%
\right. ,
\end{equation}%
and%
\begin{equation}
\left\langle j\right. \left\vert \overline{\psi }_{k}^{-}\right\rangle =%
\begin{array}{cc}
\sin \left( kj\right) , & \left( \left\vert j\right\vert \geqslant 1\right)%
\end{array}%
.
\end{equation}

The following equation can readily be confirmed:
\begin{equation}
\left\langle \overline{\psi }_{k}^{\lambda }\right. \left\vert \psi
_{k^{\prime }}^{\lambda ^{\prime }}\right\rangle =C_{k}^{\lambda }\delta
_{\lambda \lambda ^{\prime }}\delta _{kk^{\prime }},  \label{biorthonormal}
\end{equation}%
where $C_{k}^{\lambda }$\ is a bounded real function. This indicates that
the wave functions can always be renormalized to achieve a biorthogonal set,
except when $k=\pi /2$\ and\ $2\kappa \gamma =\pm g^{2}$, which implies the
collapse of the biorthonormal relation in Eq. (\ref{biorthonormal}).

\section{Spectral singularity}

\label{Spectral singularity}

In this section, we explore the scattering of the system. We show the
dynamic features at the spectral singularity~\cite{PRL3}. First, we consider
the steady-state solution of the system at the point $\left( k_{\mathrm{c}%
},\gamma _{\mathrm{c}}\right) $ with $k_{\mathrm{c}}=\pi /2$, where the wave
packet has the fast velocity $2\kappa$ and propagates without spreading. The
critical gain (loss) rate is
\begin{equation}
\gamma _{\mathrm{c}}=\sigma g^{2}/(2\kappa ),  \label{SS cond}
\end{equation}%
where $\sigma =1$\ ($\sigma =-1$) is for the gain (loss) case. When $\gamma
=\gamma _{\mathrm{c}}$, the following equations of\ scattering solution%
\begin{equation}
\left\langle j\right. \left\vert \psi _{\mathrm{c}}^{+}\right\rangle
=\left\{
\begin{array}{cc}
e^{-i\sigma \pi j/2}, & \left( j\leqslant -1\right) \\
e^{i\sigma \pi j/2}, & \left( j\geqslant 1\right)%
\end{array}%
\right. ,
\end{equation}%
and%
\begin{equation}
\left\langle j\right. \left\vert \overline{\psi }_{\mathrm{c}%
}^{+}\right\rangle =\left\{
\begin{array}{cc}
e^{i\sigma \pi j/2}, & \left( j\leqslant -1\right) \\
e^{-i\sigma \pi j/2}, & \left( j\geqslant 1\right)%
\end{array}%
\right. ,
\end{equation}%
have clear physical implications: The non-Hermitian scattering center acts
as a drain ($\sigma =1$) or source ($\sigma =-1$) of a $k=\pi /2$ plane
wave. To represent the system at spectral singularities according to the
transmission and reflection coefficients, we set a Jost solution for input
wave $k=\pi /2$:
\begin{equation}
\left\langle j\right. \left\vert \psi _{k}\right\rangle =\left\{
\begin{array}{cc}
e^{ikj}+r_{k}e^{-ikj}, & \left( j\leqslant -1\right) \\
t_{k}e^{ikj}, & \left( j\geqslant 1\right)%
\end{array}%
\right. ,
\end{equation}%
where the transmission and reflection amplitudes are obtained as%
\begin{equation}
r_{k}=-g^{2}/\eta _{k},\text{ }t_{k}=r_{k}+1.
\end{equation}%
This indicates that both $t_{k}$\ and $r_{k}$ tend to infinity at the
spectral singularity as $\eta _{k}=0$, therefore conforming the theorem
proved in Ref.~\cite{PRL3}.

\section{Wave emission}

\label{Emission} This section addressed implications of the spectral
singularity from the perspective of wave packet dynamics. In practice, we
consider the time evolution of a Gaussian wave packet with central vector $k$
to reflect the plane wave with vector $k_{\mathrm{c}}$~\cite{Kim}. The
Gaussian wave packet is%
\begin{equation}
\Phi \left( j,0\right) =\Omega ^{-1/2}e^{-\alpha ^{2}\left( j-N_{\mathrm{c}%
}\right) ^{2}/2}e^{-ikj},
\end{equation}%
where $N_{\mathrm{c}}$ is the Gaussian wave center and $\Omega =\sqrt{\pi }%
/\alpha $\ is the renormalization factor. This factor ensures the Dirac
probability of the initial state is unity. The FWHM of the Gaussian wave
packet is $\Delta =2\sqrt{2\ln 2}/\alpha $. The velocity is $2\kappa \sin
(k) $ in a uniform chain system with coupling $\kappa$.

Hamiltonian $H$ shown in Fig. \ref{fig1}(b) can be further reduced by
symmetric and anti-symmetric combination of sites $j<0$ and $j>0$, i.e., $%
a_{l_{\pm }}=(a_{-l}\pm a_{l})/\sqrt{2}$, as shown in Fig.~\ref{fig1}(c).
The system is an $N+1$-site coupled resonator array, which consists of a
uniform coupled resonator with $N=800$ and the coupling strength is $\kappa $%
. A resonator with additional loss or gain is coupled at the second last end
of the uniform array at the strength $\sqrt{2}\kappa $.

\begin{figure}[tb]
\includegraphics[bb=0 0 350 330, width=7 cm, clip]{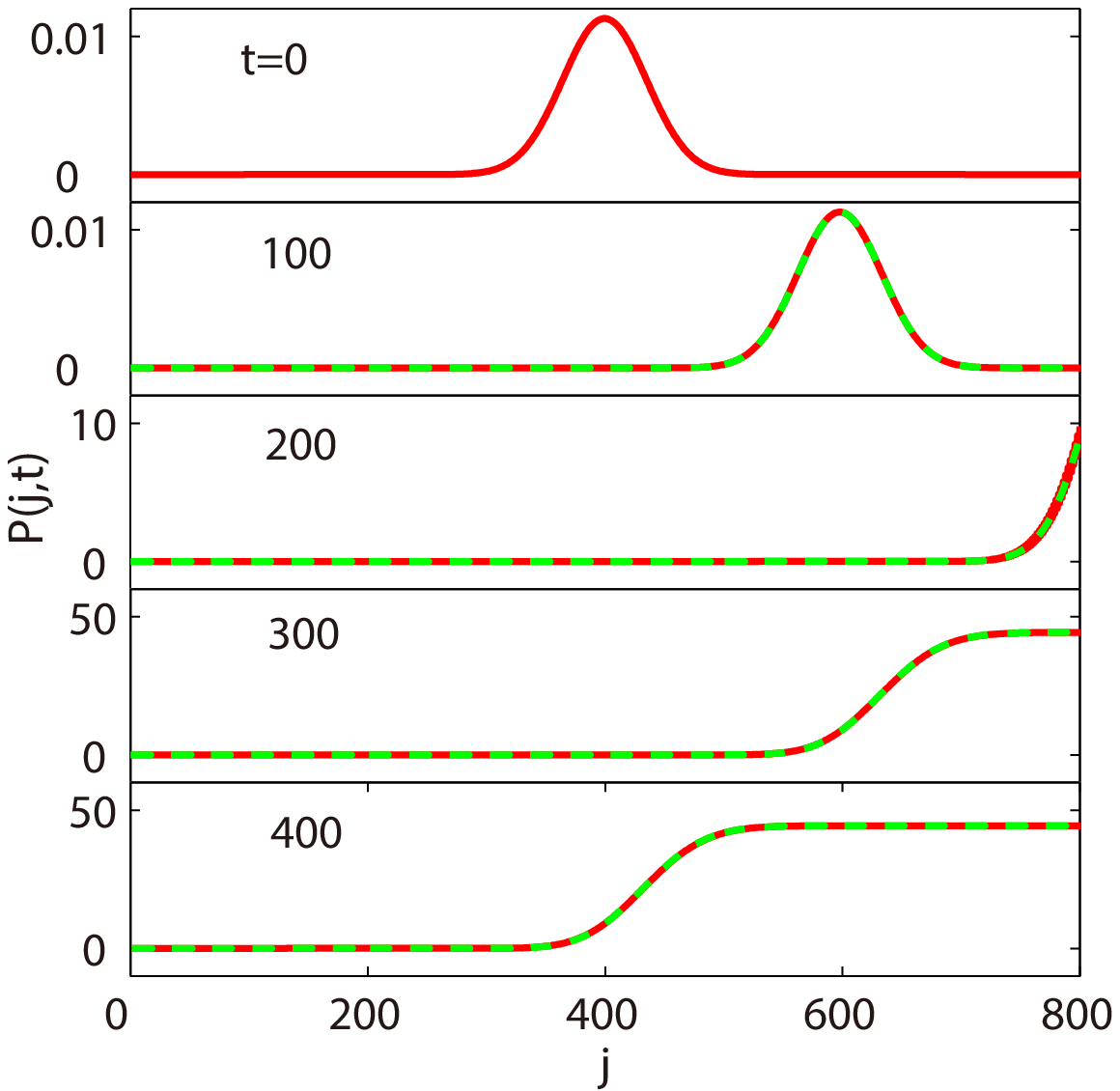}
\caption{(Color online) Scattering process of an incident photon wave packet
by the non-Hermitian scattering center. The probability distributions $P(j,t) $ for several instants were obtained according to the time evolution
under the systems of the Hamiltonian $H_{0}$ with $\protect\kappa =g=1$, and
$\protect\gamma =\protect\gamma _{\text{c}}=\protect\kappa/2$ at the
spectral singularity. The incident wave packet values were $k_{\text{c}}=\protect\pi/2 $, $N=800$, $N_{\text{c}}=400$ and $\protect\alpha =0.02$.}
\label{fig2}
\end{figure}

For $\gamma =\gamma _{\mathrm{c}}>0$, the resonator at the end is a gain
resonator. We consider a Gaussian wave packet with $\alpha =0.02$ and $k=\pi
/2$, that was centered at $N_{\mathrm{c}}=400$ at time $t=0$. The Gaussian
wave packet probability out of region $[N_{\mathrm{c}}-\Delta $, $N_{\mathrm{%
c}}+\Delta ]$ vanishes approximately and the probability within the $2\Delta
$ region around $N_{\mathrm{c}}$ is over $99.9\%$. As shown in Fig.~\ref%
{fig2}, the profiles of the Gaussian wave at different times are plotted.
The wave packet dynamics exhibited a persistent wave emission from the gain
resonator, which reflected the steady-state plane wave emission solution of
the system at the spectral singularities. The wave packet was
shape-preserving and propagated toward the gain resonator at the end of the
resonator chain array. At approximate time $t_{0}=(N-N_{\mathrm{c}}-\Delta
)/(2\kappa )\approx 141/\kappa $, the wave packet head reached the gain
resonator and the wave packet probability started increasing. The wave
packet was reflected as a concomitant amplification and finally formed a
platform with the probability height $h$ after the time $t_{0}+\Delta
/\kappa $, characterized by a Gaussian error function that previously found
in systems with complex periodic potentials~\cite{HeissArxiv,PRA2}. The details are demonstrated
in Sec. \ref{PT} and calculated in Appendix \ref{Hsolution} and \ref%
{GWPdynamics}. At the instant $t\gtrsim t_{0}+\Delta /\kappa $, the state $%
\left\vert \Phi \left( j,0\right) \right\rangle $ evolves to $\left\vert
\Phi \left( j,t\right) \right\rangle $ and the wave function has the form of
\begin{equation}
\Phi \left( j,t\right) =-\frac{\sqrt{h}}{2}\{1+\mathrm{erf}[\frac{2^{3/4}}{%
\Delta }\left( j-N_{t}\right) ]\}e^{-\frac{i\pi j}{2}},  \label{Phi_t}
\end{equation}%
where $N_{t}=2(N+1)-N_{\mathrm{c}}-2\kappa t$ represents the Gaussian wave
packet center after reflection for $\gamma =0$, $h=2(\gamma _{\mathrm{c}%
}/\kappa )^{2}\sqrt{\pi }/\alpha $ shows the platform wave emission
probability spreading out from the gain resonator to infinity. The wave
front of $\Phi \left( j,t\right) $ has the velocity $2\kappa $.

\begin{figure}[tbp]
\includegraphics[bb=0 0 470 235, width=8.6 cm, clip]{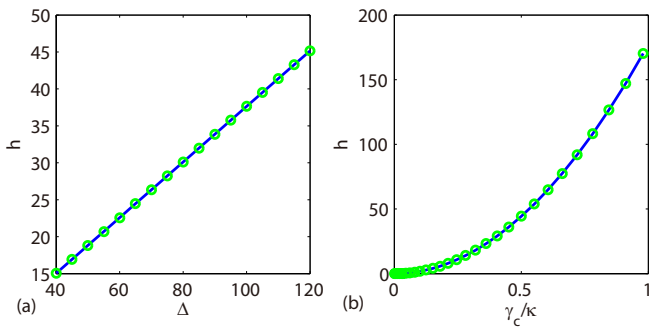}
\caption{(Color online) Scaling of the emission wave probability platform
height $h$ as a function of (a) the FWHM $\Delta$ of a Gaussian wave packet
at $g/\protect\kappa=1$, and as a function of (b) the resonator gain $\protect\gamma_{\text{c}}$ at the spectral singularities at $\protect\alpha=0.02$. The green circles were obtained through numerical simulation and the
blue lines were plots from Eq.~(\protect\ref{h}).} \label{fig3}
\end{figure}

Figure~\ref{fig3}(a) shows the scaling of wave-emission platform height $h$
as a function of the FWHM $\Delta $; in Fig.~\ref{fig3}(b), the scaling of $%
h $ as a function of the resonator gain $\gamma $ is shown.%
\begin{equation}
h=\sqrt{\frac{\pi }{2\ln 2}}(\gamma _{\mathrm{c}}/\kappa )^{2}\Delta ,
\label{h}
\end{equation}%
the probability height $h$ linearly depends on the FWHM $\Delta $, and
quadratically depends on the resonator gain $\gamma _{\mathrm{c}}$ at the
spectral singularity.

\section{Wave absorption}

\label{Absorption} 
\begin{figure}[tb]
\includegraphics[ bb=0 0 350 330, width=7 cm, clip]{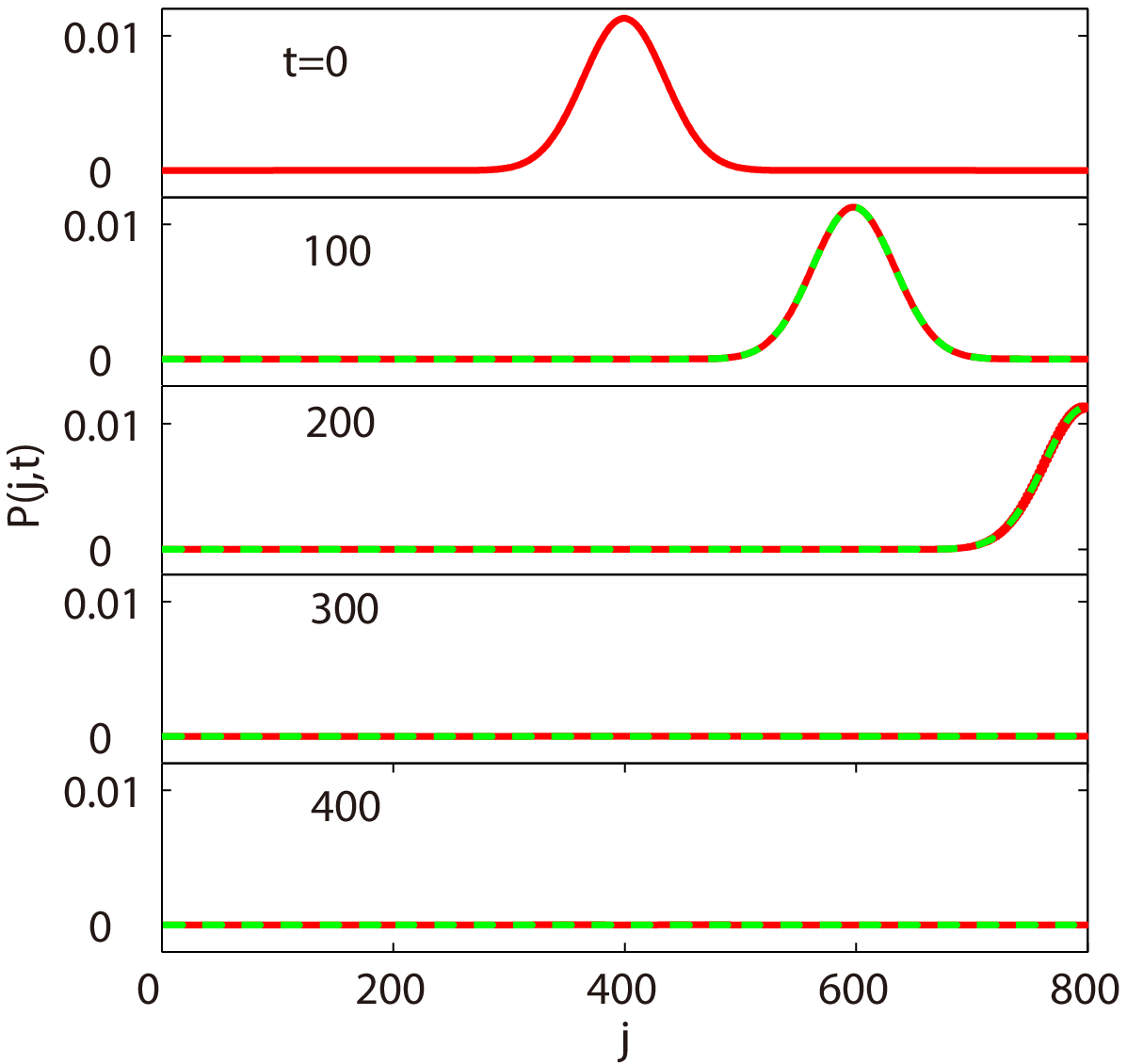}
\caption{(Color online) Scattering process of an incident photon wave packet
by the non-Hermitian scattering center. The probability distributions $P(j,t) $ for several instants were obtained according to the time evolution
under the systems of the Hamiltonian $H_{0}$ with $\protect\kappa =g=1$, and
$\protect\gamma =-\protect\gamma _{\mathrm{c}}=-0.5$ at the spectral singularity. The
incident wave packet values were $k_{\mathrm{c}}=\protect\pi/2$, $N=800$, $N_{\mathrm{c}}=400$
and $\protect\alpha =0.02$.} \label{fig4}
\end{figure}

For $\gamma =-\gamma _{\mathrm{c}}<0$, the time evolution
of $\left\vert \Phi \left( j,0\right) \right\rangle $ differs from the wave
emission case ($\gamma =\gamma _{\mathrm{c}}>0$). The state $\left\vert \Phi
\left( j,0\right) \right\rangle $ will be absorbed at the lossy site. As
shown in Fig.~\ref{fig4}, a Gaussian wave packets with $k=\pi /2$, centered
at $N_{\mathrm{c}}=400$ at time $t=0$, moves toward the lossy resonator at
the end and is absorbed without reflection because of the dissipation. At
approximate time $(N-N_{\mathrm{c}}-\Delta )/(2\kappa )\approx 141/\kappa $,
the head of the Gaussian wave packet with $\alpha =0.02 $ reaches the lossy
resonator $N$, and the arrived component is completely absorbed without
reflection. At time $t_{0}=(N-N_{\mathrm{c}})/(2\kappa )=200/\kappa $, the
wave packet center arrives at the lossy resonator and half of the wave
packet has been perfectly absorbed. After the wave packet moves forward for
an extra distance of $\Delta $ with an additional time $\Delta /(2\kappa )$
cost [i.e., at time $t_{0}+\Delta /(2\kappa )$], the tail of the wave packet
is perfectly absorbed. The absorption demonstrated in the configuration
shown in Fig.~\ref{fig1}(c) corresponds to the coherent perfect absorption
in the configuration of Fig.~\ref{fig1}(b). In a coherent perfect
absorption, the appropriate amplitudes and phases of the two counter
propagating incidences are perfectly absorbed~\cite{CPA,CPAexp,Science2011}.
The wave packet in the configuration depicted in Fig.~\ref{fig1}(c) implies
that the two wave packets with the same profile and wave vector that are
located at symmetric positions near the lossy resonator $j=0$ in the
configuration shown in Fig.~\ref{fig1}(b), move in opposite directions and
are perfectly absorbed at the lossy resonator $j=0$.

The time evolution of the initial state $\left\vert \Phi \left( j,0\right)
\right\rangle $ is simply a moving Gaussian wave with velocity $2\kappa$,
the center of which is $N_{t}=N_{\mathrm{c}}+2\kappa t$ before reaching the
loss site,
\begin{equation}
\Phi \left( j,t\right) =\Omega ^{-1/2}e^{-\alpha ^{2}\left( j-N_{t}\right)
^{2}/2}e^{-i\frac{\pi }{2}j},
\end{equation}%
the Gaussian wave packet vanishes out of region $[N_{t}-\Delta ,N_{t}+\Delta
]$ because of the exponential decay of the wave packet tails. When the head
of the wave packet reaches the lossy resonator and $N_{t}+\Delta >N$, the
region $[N,N_{t}+\Delta ]$ is perfectly absorbed but $[N_{t}-\Delta ,N]$
remains. After the tail of the wave packet passes the end of the resonator
array when $N_{t}-\Delta >N$, the Gaussian wave packet is completely
absorbed.

\section{$\mathcal{PT}$-symmetric system with gain and loss}

\label{PT}

\begin{figure}[tb]
\includegraphics[bb=0 0 450 50, width=8 cm, clip]{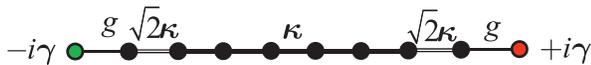}
\caption{(Color online) Schematic configuration of the $\mathcal{PT}$-symmetric system combining by two semi-infinite scattering systems cut at the
leads.} \label{fig5}
\end{figure}

In this section, we consider a $\mathcal{PT}$-symmetric system with finite
sites (Fig.~\ref{fig5}). The finite $\mathcal{PT}$-symmetric system is
formed by connecting a finite gain structure and a finite loss structure.
The finite gain (loss) structure is cut at the lead of the semi-infinite
chain array for wave emission (perfect absorption) as shown in Fig.~\ref%
{fig1}(c). The $\mathcal{PT}$-symmetric Hamiltonian can be described by a
one-dimensional finite chain array, the ends of which are coupled to two
resonators with balanced gain and loss. The Hamiltonian is%
\begin{eqnarray}
H_{\mathcal{PT}} &=&-\kappa (\sum_{l=3}^{N-1}a_{l}^{\dag }a_{l+1}+\sqrt{2}%
a_{2}^{\dag }a_{3}+\sqrt{2}a_{N}^{\dag }a_{N+1}+\text{\textrm{H.c.}})  \notag
\\
&&-g(a_{1}^{\dag }a_{2}+a_{N+1}^{\dag }a_{N+2}+\text{\textrm{H.c.}})  \notag
\\
&&-i\gamma a_{1}^{\dag }a_{1}+i\gamma a_{N+2}^{\dag }a_{N+2}.  \label{H_PT}
\end{eqnarray}%
We consider the $\mathcal{PT}$-symmetric system as a finite gain
structure coupled to its $\mathcal{PT}$-symmetric loss countpart. Thus, the $%
\mathcal{PT}$-symmetric system has even site $N$. The exceptional point is
at $\gamma _{\mathrm{c}}=g^{2}/2\kappa $\ as demonstrated in detail in
Appendix \ref{Hsolution}. In weak gain or loss region $|\gamma |\leqslant
\gamma _{\mathrm{c}}$, the system has two pairs of bound states with complex
eigenvalues; in strong gain or loss region $|\gamma |>\gamma _{\mathrm{c}}$,
one more pair of bound states emerge with pure imaginary eigenvalues. The
system is nondiagonalizable consisting of a $2\times 2$ Jordan block. The
coalesced state is a plane wave state that emits from the gain resonator to
the loss resonators. The plane wave state is the coalescence of the wave
emission state and the absorption state at the spectral singularity of the
semi-infinite systems, with gain and loss resonator at the ends.

\begin{figure}[tbp]
\includegraphics[ bb=0 0 410 417, width=6 cm, clip]{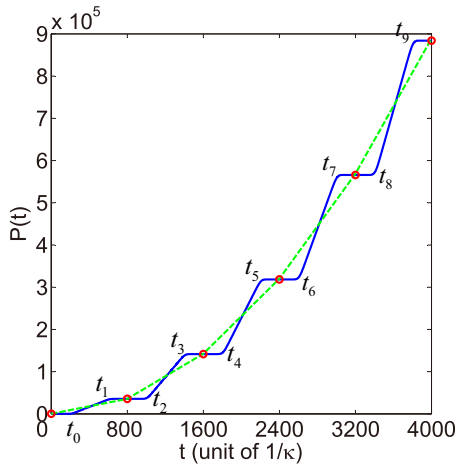}
\caption{(Color online) Time evolution probability of Gaussian wave in a $\mathcal{PT}$-symmetric finite system (blue curve). Red circles show the
probability when the wave front is at the center of the $\mathcal{PT}$-symmetric system. The green curve indicates the quadratic increasing of
wave packet probability in a long time scale. $t_n$ marks the time of the
wave front reflected at the ends resonator with gain and loss. The
parameters are $g=\protect\kappa=1$, $\protect\alpha=0.02$, $k=\protect\pi/2$, $N=800$, $N_{\rm{c}}=400$.}
\label{fig6}
\end{figure}

In Fig.~\ref{fig6}, we show a Gaussian wave with $k=\pi /2$, centered at $N_{%
\mathrm{c}}=400$, moving with velocity $2\kappa $ toward the gain resonator.
In order to simulate wave packet dynamics, we omit the significantly small
probabilities of wave packet on the two pairs of bound states, which are far
away from the wave packet center in the momentum space. At the time $%
t_{0}=200/\kappa $ (reaching the system ends for the first time). The wave
packet probability starts increasing because it reaches the gain resonator.
The emission wave at the gain resonator leads to a linear growth after wave
packet reflection. The reflection at the gain resonator realizes a plane
wave emission process as discussed in the preceding section. The emission
wave still has a velocity of $2\kappa $ and at time $t_{1}=600/\kappa $, the
wave reaches the loss point and is absorbed. After the wave is reflected at
the loss resonator, the waves that are emitted and absorbed are balanced and
the probability remains constant until the wave front reaches the gain
resonator again at $t_{2}=1000/\kappa $. In the time interval $[t_{1},t_{2}]$%
, the wave emission at the gain resonator and wave absorption at the loss
resonator are balanced. The wave increases again at the time interval $%
[t_{2},t_{3}]$. After the wave front touches the loss resonator at $%
t_{3}=1400/\kappa $, the dynamics of the left propagating wave with constant
probability and the right propagating wave with a linear time dependent
probability growth repeat as described. $t_{n}=t_{0}+mN/(2\kappa)$ ($m\in
\mathbb{Z}$) is the time wave reflection happens at the resonators with gain
and loss. The dynamics of $2\times 2$ Jordan blocks for dimer systems have
been thoroughly investigated. The probability in the dimer exhibits an
increasing power law with the highest order being a quadratic function~\cite%
{GraefePRL,GraefePRA,MoiseyevEP11}. The increasing quadratic is associated
with the $2\times 2$ Jordan block. For a higher order Jordan block (e.g., a $%
3\times 3$ Jordan block), the highest order of the probability increase is
higher (quartic function). We note that in the $\mathcal{PT}$-symmetric
system, the wave packet probability increases quadratically in a large time
scale
\begin{equation}
P\left( t\right) \approx 1+(h/N)(\kappa t)^{2},
\end{equation}%
as denoted by the dashed green line shown in Fig.~\ref{fig6}, where the
coefficient $h/N\approx 1/18$. The red circles in Fig.~\ref{fig6} indicate
the probabilities of the wave at the chain center at time $mN/\kappa $ ($%
m\in \mathbb{Z}$). The quadratic probability increase reflects that the
system is at the exceptional point. The Hamiltonian system consists of a $%
2\times 2$ Jordan block with two coalescence states. Thus, we link the
spectral singularities of a semi-infinite scattering system with the
exceptional point of a finite $\mathcal{PT}$-symmetric system.

In the Appendix, we calculate the dynamics of a Gaussian wave packet in the $%
\mathcal{PT}$-symetric system. In a large system that the wave packet width
far less than the system size ($\Delta \ll N$), the wave packet is localized
without spreading in the dynamical process before the wave front reach the
system ends. The system has no long-range interactions and is described by a
tight-binding model. The loss far away from the gain in the system hardly
affects the dynamics of a wave packet that close to the area near the gain
site, therefore, the dynamics in the $\mathcal{PT}$-symmetric system is
approximately the same as that in the corresponding wave emission system.

\section{Near the spectral singularities}

\begin{figure}[tbp]
\includegraphics[ bb=80 185 537 590, width=4.0 cm, clip]{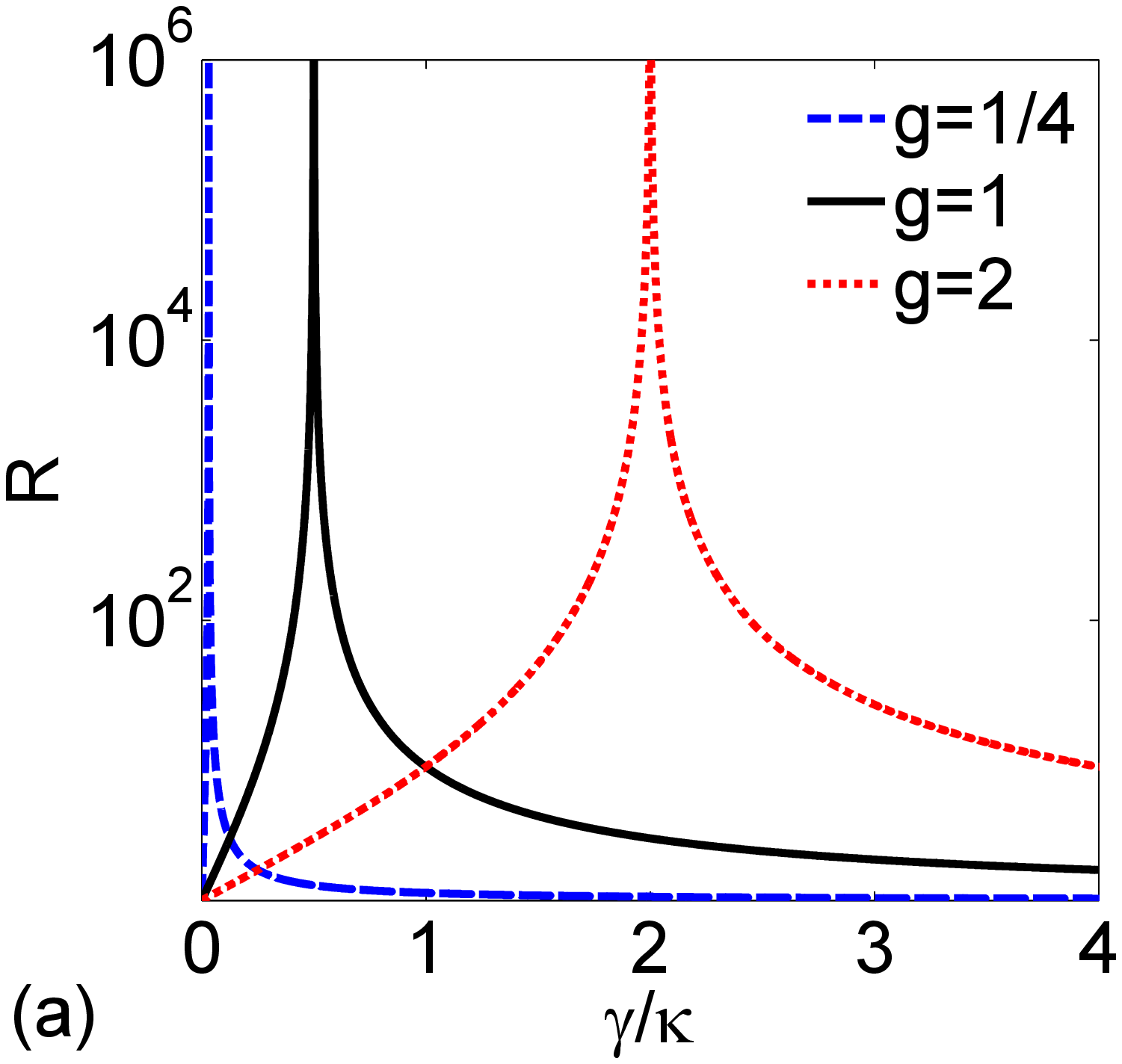} %
\includegraphics[ bb=80 185 537 590, width=4.0 cm, clip]{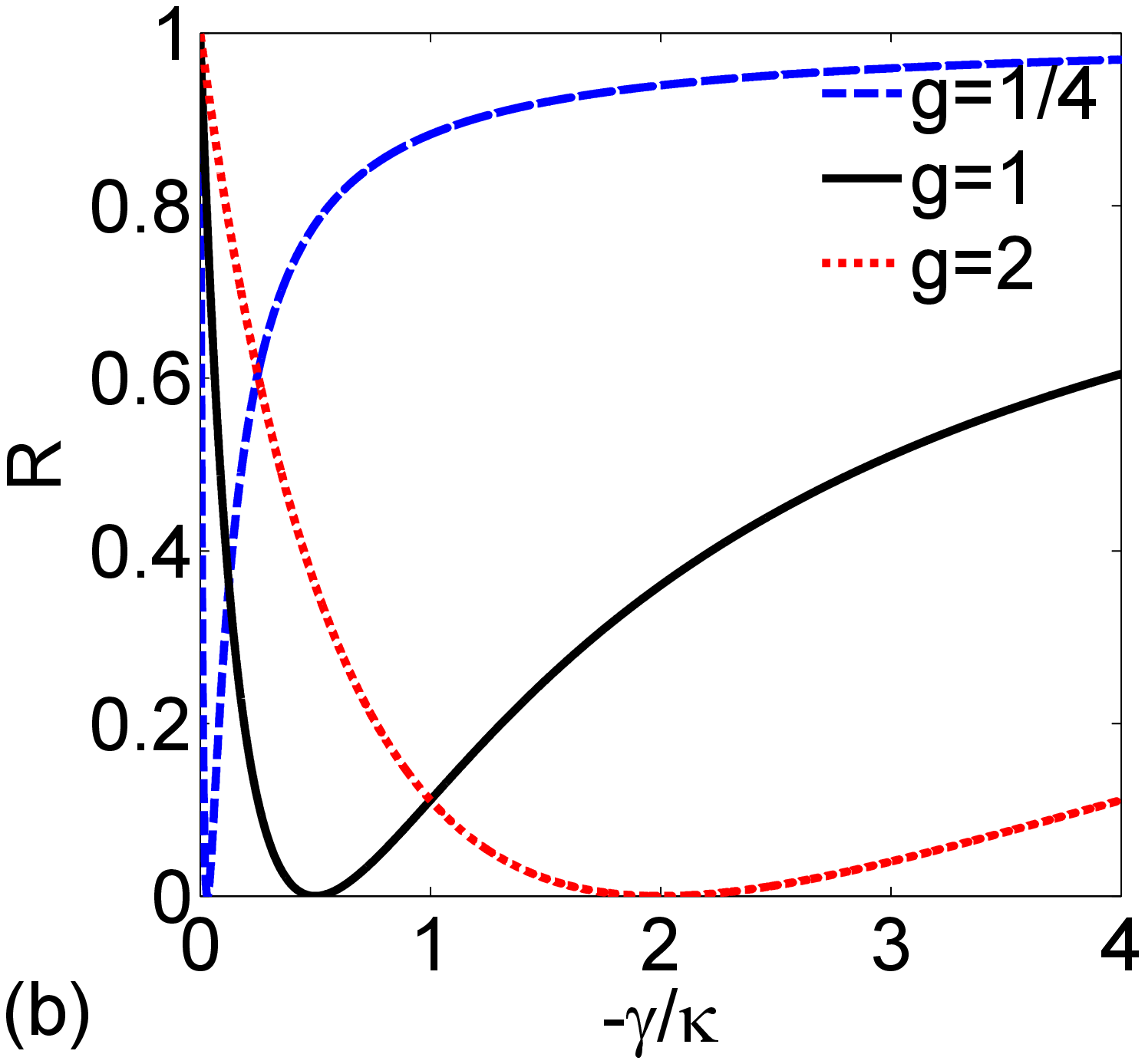} %
\includegraphics[ bb=20 180 480 605, width=4.0 cm, clip]{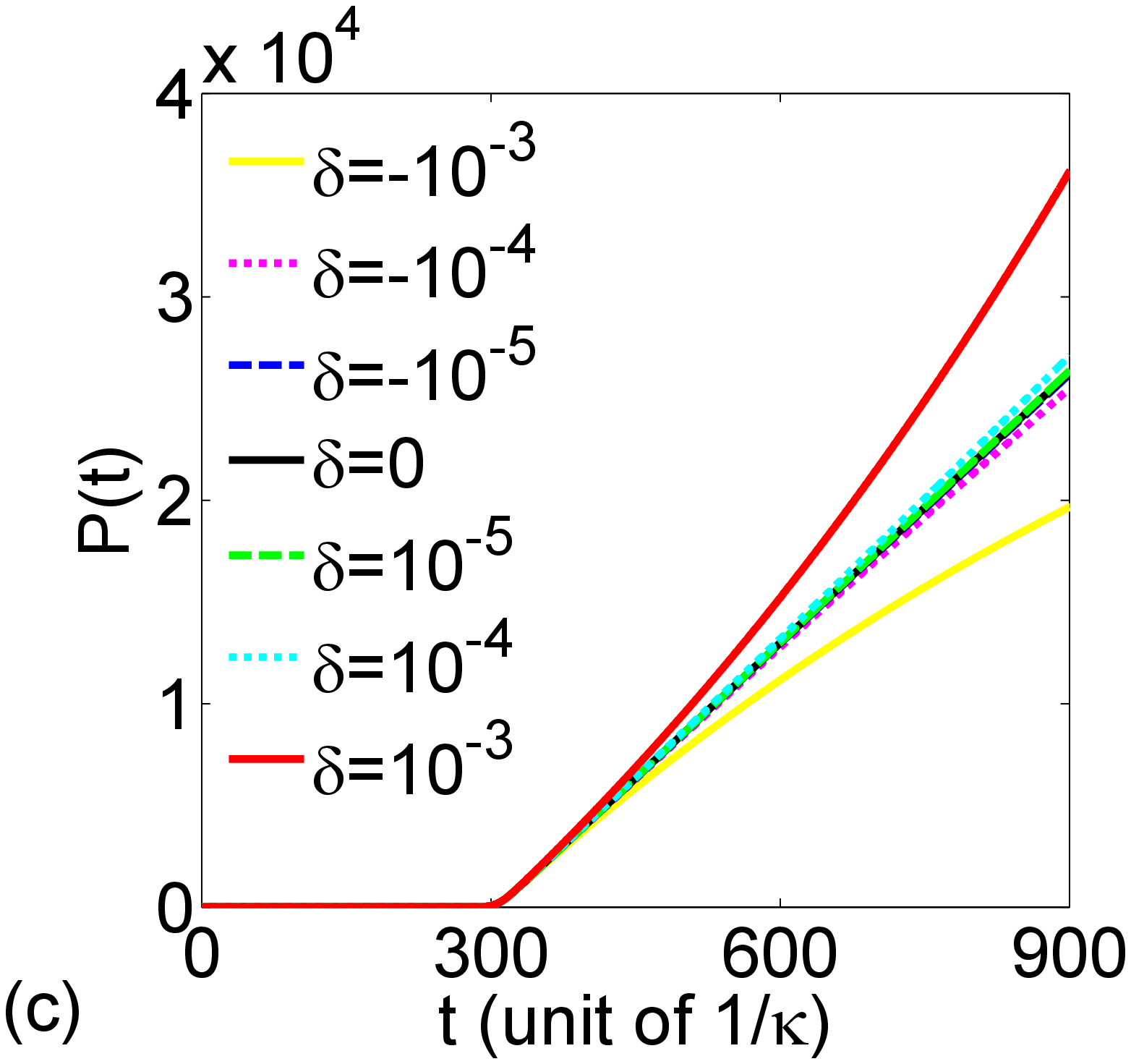} %
\includegraphics[ bb=20 180 480 605, width=4.0 cm, clip]{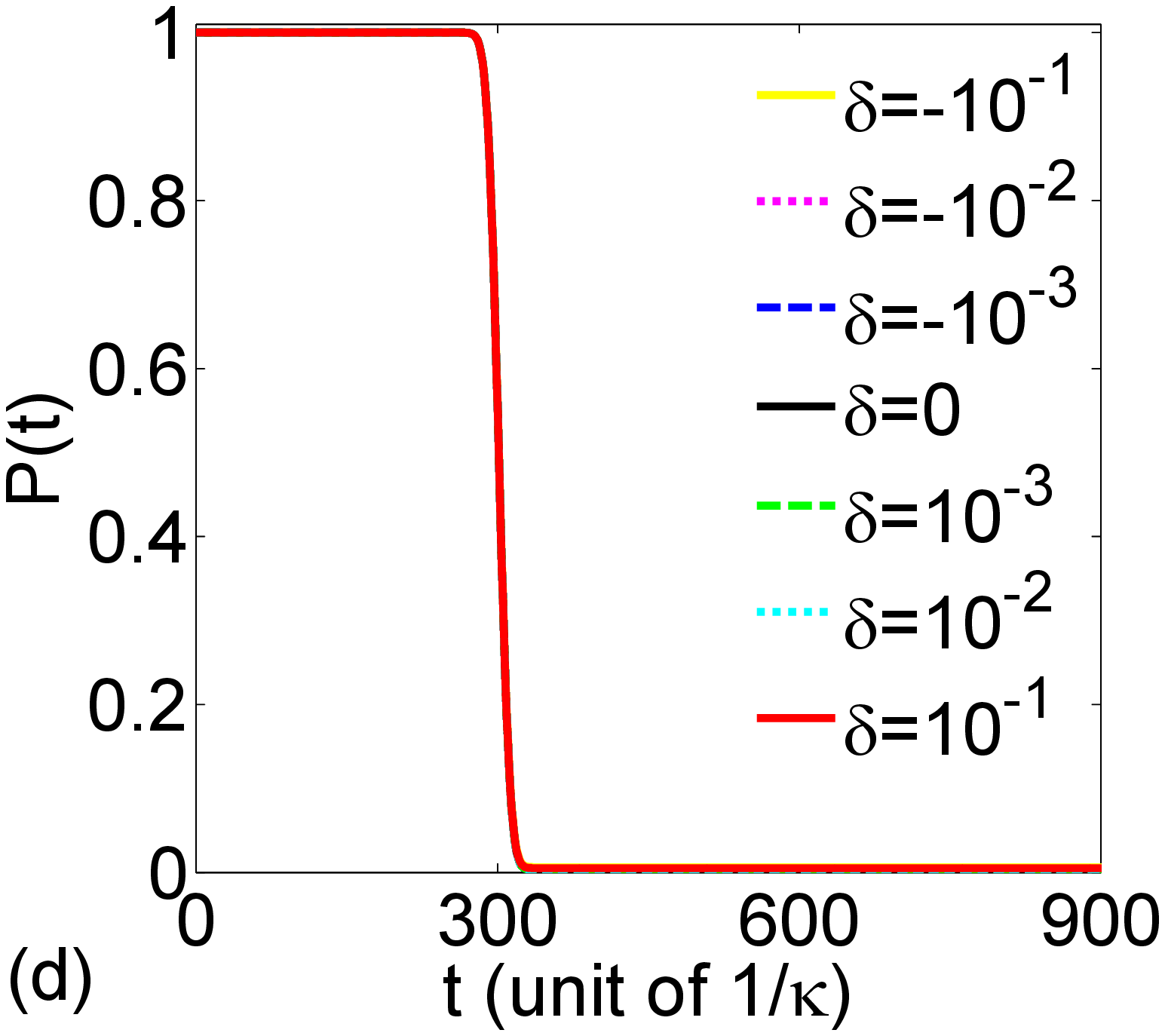} %
\includegraphics[ bb=20 180 480 605, width=4.0 cm, clip]{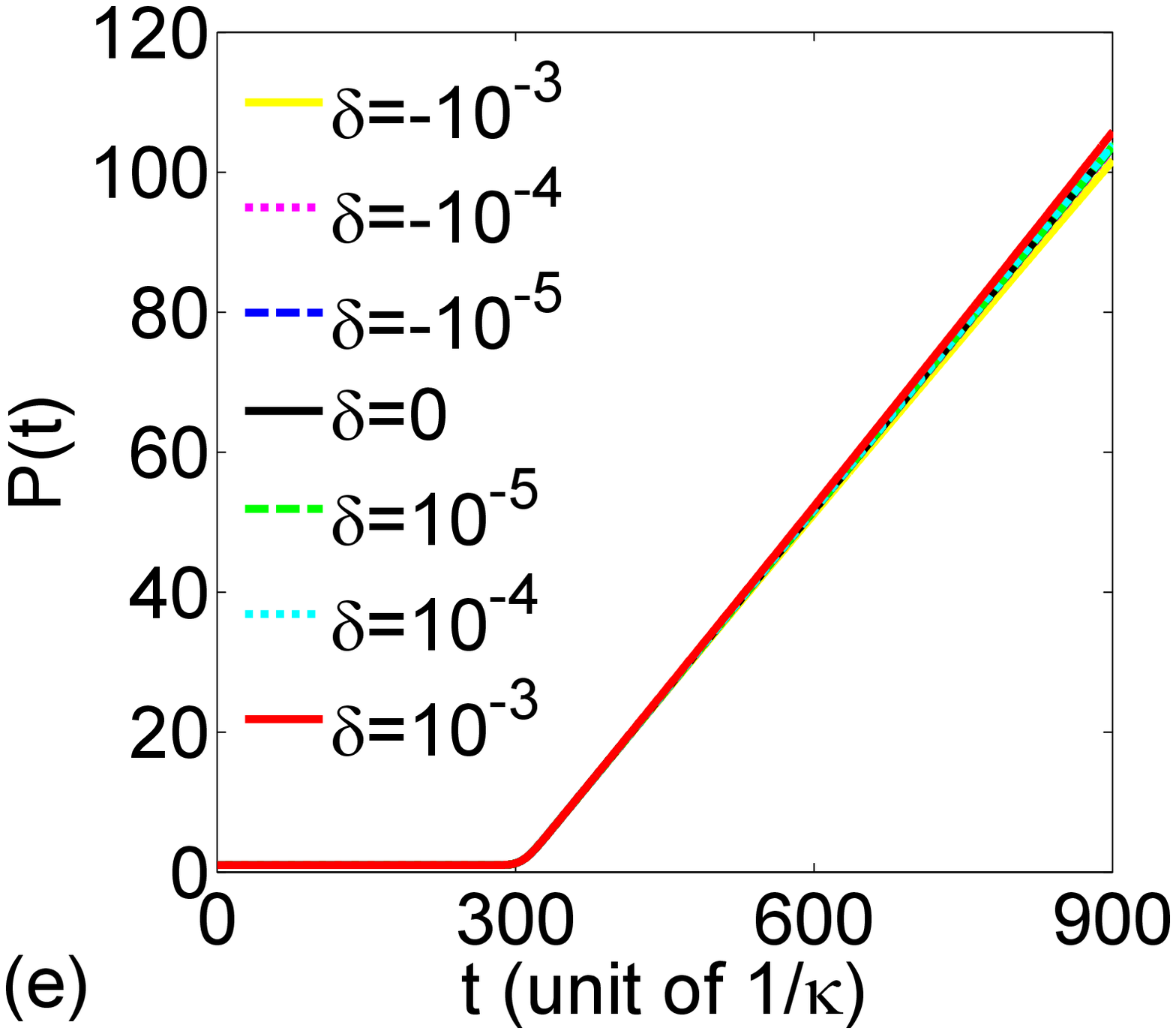} %
\includegraphics[ bb=20 180 480 605, width=4.0 cm, clip]{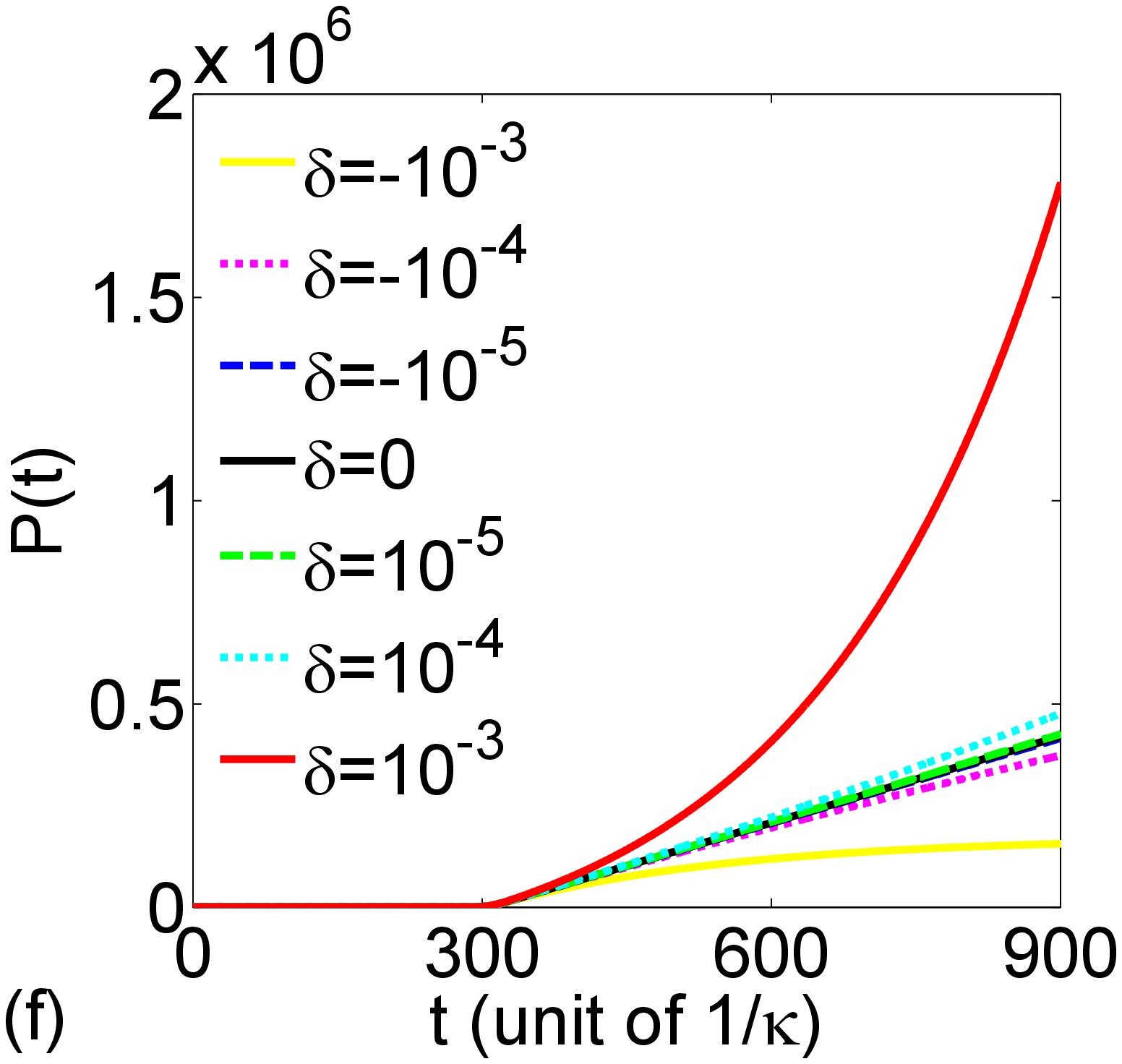}
\caption{(Color online) The reflection probability and the dynamics near the spectral singularities for the configuration Fig.~\ref{fig1}(c). The reflection $R(\gamma)$ for gain (a) and loss (b). (c-f) The curves correspond to systems with different deviations $\delta$ in the gain (c, e, f) or loss (d).  The
parameters are $\kappa=1$, $\protect\alpha=0.04$, $k=\protect\pi/2$, $N=800$, $N_{\rm{c}}=200$; $g=1$ in (c, d), $g=1/4$ in (e), $g=2$ in (f).}
\label{fig7}
\end{figure}

The characteristic dynamics at and near the system exceptional points was
investigated in several previous works~\cite{MoiseyevEP11,HeissEP10}. The
polynomial dynamical features at the exceptional points were also found near
the exceptional points in a narrow region. Here in this section, we show the
dynamical behaviors near the spectral singularities. The reflection
coefficient for $k=\pi /2$ wave in configuration Fig.~\ref{fig1}(c) is $%
r=(2\gamma +g^{2})/(2\gamma -g^{2})$. For $\gamma >0$\ ($\gamma <0$), the
reflection is larger (smaller) than unity, which is attributed to the gain
(loss) at the system end. The reflection probability $R(\gamma )=|r|^{2}$\
is shown in Fig.~\ref{fig7}(a,b). The reflection coefficient diverges at the
spectral singularities for system with gain, and changes dramatically near
the spectral singularities [Fig.~\ref{fig7}(a)]. For system with loss, the
reflection coefficient is the inverse, i.e., $R(-\gamma )=R(\gamma )^{-1}$,
the reflection at the spectral singularities is zero. Contrary to the system
with gain, the reflection is insensitive as the loss $\gamma $\ near the
spectral singularities, the reflection probability is close to zero in a
wide region near the spectral singularities, in particular at large $\gamma
_{\mathrm{c}}$. [Fig.~\ref{fig7}(b)].

In order to investigate the dynamics near the spectral singularities, we
define the gain or loss $\gamma $ deviated from the spectral singularities
by a deviation parameter $\delta $, as
\begin{equation}
\gamma =\gamma _{\mathrm{c}}(1+\delta ).
\end{equation}%
The dynamical difference is enlarged as time increasing for $\gamma >0$. We
focus on a short time internal within the wave packet reaching the gain site
for the second time as shown in Fig.~\ref{fig7}(c-f). In Fig.~\ref{fig7}%
(c,d), we show the dynamical time evolution of the probability of an
incident wave packet at $g=\kappa =1$. At about $t=300/\kappa $\ in Fig.~\ref%
{fig7}(c), the wave emission starts, the probability increases linearly at
the spectral singularity. For the gain $\gamma $\ slightly deviates from the
critical gain $\gamma _{\mathrm{c}}$\ at about $\delta =10^{-4}$, the time
evolution keeps close with the linear increasing in a time interval before
the wave packet reaching the gain site once more after reflection. We
consider the probability difference ($|1-P_{\gamma }/P_{\gamma _{\mathrm{c}%
}}|$) at $t=900/\kappa $. The probability difference remains within $3\%$\
for $|\delta |\leqslant 10^{-4}$, the probability difference for $|\delta
|=10^{-3}$\ is remarkable, reaches about $40\%$. In Fig.~\ref{fig7}(d), the
perfect absorption in a loss system remains good for large deviation. From
the expression of reflection $R(\gamma)$, the probability without absorption
is around $1\%$\ for loss deviation as large as $|\delta |=10^{-1}$. After
the wave packet reached the loss site for the first time (around $%
t=300/\kappa $), the wave packet is almost fully absorbed.

In Fig.~\ref{fig7}(e,f), we show the time evolution of probability at
different spectral singularities. The spectral singularities at different
critical gain $\gamma _{\mathrm{c}}$\ varies as the coupling $g$. Strong
coupling $g$\ results in large gain $\gamma _{\mathrm{c}}$. The probability
difference in the time evolution process is gain dependent, the significant
probability difference is caused by the large gain. As the critical gain
reduced to $\gamma _{\mathrm{c}}=\kappa /32$\ at $g=\kappa /4$, the
probability difference decreases to $2\%$\ for $\delta =10^{-3}$\ [Fig.~\ref%
{fig7}(e)]; as the critical gain increased to $\gamma _{\mathrm{c}}=2\kappa $%
\ at $g=2\kappa $, the probability difference increases to $320\%$\ for $%
\delta =10^{-3}$\ [Fig.~\ref{fig7}(f)]. Moreover, through comparison, we
note that the probability difference is smaller for negative $-|\delta |$\
than that of positive $|\delta |$, which also indicates that large gain
leading to more probability difference. For a system with loss, the
absorption is insensitive near the spectral singularities in a wide region,
in particular for systems with large loss.

\begin{figure}[tbp]
\includegraphics[bb=0 0 575 355, width=8.7 cm, clip]{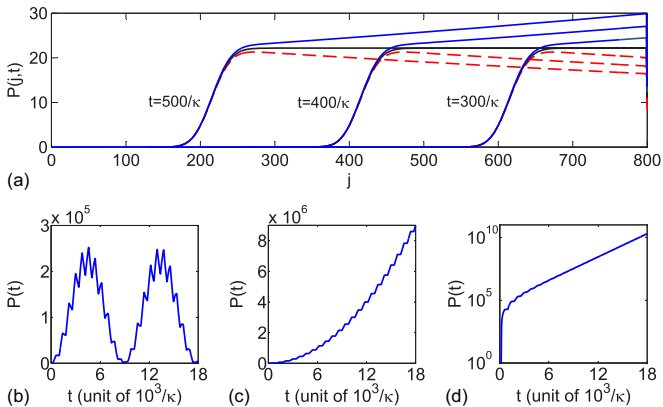}
\caption{(Color online) The probabilities of Gaussian wave packet in the
systems near the spectral singularities. $g=\kappa=1$,
$\protect\alpha=0.04$, $k=\protect\pi/2$, $N=800$, $N_{\rm{c}}=400$. (a) $|\delta|=0,  10^{-3}$ for the
scattering systems; (b) $\delta=-10^{-4}$, (c) $\delta=0$, (d)
$\delta=10^{-4}$ for the $\mathcal{PT}$-symmetric systems.} \label{fig8}
\end{figure}

In Fig.~\ref{fig8}(a), the time evolution profiles of Gaussian wave packet
in the scattering system as illustrated in Fig.~\ref{fig1}(c) are shown for $%
|\delta |=0,10^{-3}$. The red dashed (blue solid) lines are for $\delta
=-10^{-3}$ ($\delta =10^{-3}$), they are similar but slight smaller (larger)
than the platform wave emission at the spectral singularity (black solid
lines), the profiles deviations diminish as $|\delta |$ decrease. In Fig.~%
\ref{fig8}(b-d), the time evolution of probabilities in the $\mathcal{PT}$%
-symmetric system (Fig.~\ref{fig5}) are shown for different deviations $%
\delta $. The probability oscillates in the region $\gamma <\gamma _{\mathrm{%
c}}$ in Fig.~\ref{fig8}(b), quadratically increases at the exceptional point
in Fig.~\ref{fig8}(c), and exponentially increases in the region $\gamma
>\gamma _{\mathrm{c}}$ in Fig.~\ref{fig8}(d). The exponential increasement
is induced by the one more pair of bound states emerged with pure imaginary
eigenvalues when $\gamma >\gamma _{\mathrm{c}}$.

\section{Conclusion}

\label{Summary}We studied the spectral singularity of a semi-infinite
coupled resonator system. The system is a coupled resonator array with a
gain (lossy) resonator at the end. This system can be described by a
one-dimensional tight-binding chain. Based on the exact solution, the
critical dynamical behavior associated with the spectral singularities was
demonstrated. The dissipative resonator absorbs the incoming wave with a
matched wave vector. When the resonator loss is replaced by a gain, the
coupled resonator system remains at the spectral singularity. However, the
dynamics is a persistent wave emission instead of a perfect absorption. We
found that the wave emission can be characterized by a time dependent
Gaussian error function. The emission wave amplitude probability linearly
depends on the FWHM of the incident Gaussian wave, and quadratically depends
on the resonator gain. Combination of the scattering systems at spectral
singularities can form a finite $\mathcal{PT}$-symmetric system at
exceptional point. Thus, we linked the spectral singularity to the
exceptional point. The coalescence states are the absorption and emission
solutions at the spectral singularities in the scattering system. The
occurrence of two coalescent states indicates that the $\mathcal{PT}$%
-symmetric system has a $2\times 2$ Jordan block and that the probability
increases quadratically. Calculating the time evolution of Gaussian wave
packet with energy on resonance of the coalesced states, we revealed a
stepped progressive ascent of probability. For a long time scale, the
probability increases quadratically, which reflects the exceptional point;
for a short time scale, the plane wave emission and absorption phenomena
reflect the dynamics at the spectral singularities. The dynamical features
near the spectral singularities were investigated, the dynamical differences
in the wave packet profiles and probabilities induced by slight deviation
from the critical gain (loss) was shown, the effect bringing by the
deviation is significant (tiny) at large critical gain (loss).

\acknowledgments We acknowledge the support of CNSF (Grant Nos. 11605094 and
11374163), the Baiqing Plan of Nankai University (Grant No. ZB15006104).

\appendix

\label{Appendix}

\section{Solution of the $\mathcal{PT}$-symmetric system}

\label{Hsolution} We consider the $\mathcal{PT}$-symmetric Hamiltonian Eq. (%
\ref{H_PT}). Introducing a local unitary transformation $u$, which is
defined by $ua_{j}^{\dagger }u^{-1}=(-1)^{j}a_{j}^{\dagger }$, $%
ua_{j}u^{-1}=(-1)^{j}a_{j}$, we have $(\mathcal{P}u)H(\mathcal{P}u)^{-1}=-H$%
, i.e., $H$\ has a chiral symmetry. On the other hand, unitary
transformation $\mathcal{P}u$\ maintains the spectrum of the system. Then
the spectrum is symmetric about zero energy. From the investigation of the
zero energy solution, we obtain one zero energy state for system with odd $N$%
, and two coalesced zero energy states at the exceptional point $\gamma _{%
\mathrm{c}}=g^{2}/(2\kappa )$ for system with even $N$ (the situation we
discussed in Sec. \ref{PT}: The $\mathcal{PT}$-symmetric system is a finite
gain structure coupled to its $\mathcal{PT}$-symmetric loss structure
countpart).

In general situation, the single-particle solution for any $N$ has the form%
\begin{equation}
\left\vert \varphi _{k}\right\rangle =\sum_{l}f_{l}\left\vert l\right\rangle
,  \label{WF}
\end{equation}%
the wave function ansatz is $f_{l}=Ae^{ikl}+Be^{-ikl}$ for the uniform
region of the system ($3\leqslant l\leqslant N$). To be concise, $\sum_{l}$
represents the summation of all $l$ ranging from $1$ to $N+2$.\ The Schr\"{o}%
dinger equation can be expressed in the explicit form of%
\begin{eqnarray}
-gf_{2} &=&\left( E_{k}+i\gamma \right) f_{1}, \\
-gf_{1}-\sqrt{2}\kappa f_{3} &=&E_{k}f_{2}, \\
-\sqrt{2}\kappa f_{2}-\kappa f_{4} &=&E_{k}f_{3}, \\
-\kappa f_{N-1}-\sqrt{2}\kappa f_{N+1} &=&E_{k}f_{N}, \\
-gf_{N+2}-\sqrt{2}\kappa f_{N} &=&E_{k}f_{N+1}, \\
-gf_{N+1} &=&\left( E_{k}-i\gamma \right) f_{N+2},
\end{eqnarray}%
where $E_{k}=-2\kappa \cos k$ is the eigen energy. Substituting $%
f_{l}=Ae^{ikl}+Be^{-ikl}$ ($3\leqslant l\leqslant N$), we obtain%
\begin{widetext}
\begin{eqnarray}
-gf_{2}-\left( E_{k}+i\gamma \right) f_{1} &=&0,  \label{EQ1} \\
-gf_{1}-\sqrt{2}\kappa (Ae^{3ik}+Be^{-3ik})-E_{k}f_{2} &=&0,  \label{EQ2} \\
-\sqrt{2}\kappa f_{2}-\left( E_{k}e^{3ik}+\kappa e^{4ik}\right) A-\left(
E_{k}e^{-3ik}+\kappa e^{-4ik}\right) B &=&0,  \label{EQ3} \\
-\sqrt{2}\kappa f_{N+1}-(E_{k}e^{ikN}+\kappa e^{ik\left( N-1\right)
})A-(E_{k}e^{-ikN}+\kappa e^{-ik\left( N-1\right) })B &=&0,  \label{EQ4} \\
-gf_{N+2}-\sqrt{2}\kappa (Ae^{ikN}+Be^{-ikN})-E_{k}f_{N+1} &=&0,  \label{EQ5}
\\
-gf_{N+1}-\left( E_{k}-i\gamma \right) f_{N+2} &=&0,  \label{EQ6}
\end{eqnarray}%
after eliminating the coefficients of wave function, the wave vector $k$
satisfies%
\begin{equation}
\{ 4\kappa ^{2}[ \kappa ^{2}\sin ^{2}\left( 2k\right) +\gamma
^{2}\sin ^{2}k] -g^{4} \} \sin \left[ \left( N-1\right) k\right]
+4g^{2}\kappa ^{2}\sin \left( 2k\right) \cos \left[ \left( N-1\right) k%
\right] =0.  \label{CriticalEQ}
\end{equation}
\end{widetext}It is easy to check that $k=\pi /2$ is a solution of Eq. (\ref%
{CriticalEQ}) for arbitrary system parameters with odd $N$; but the
parameters should satisfy $\gamma =g^{2}/(2\kappa )$ for system with even $N$.

Now we consider the wave vector $k=\pi /2$. The corresponding wave function $%
\left\vert \varphi _{\pi /2}\right\rangle $ at $\gamma =g^{2}/\left( 2\kappa
\right) $ (i.e., $E_{k}=0$) under the Schr\"{o}dinger equations of
Hamiltonian $H$ yields%
\begin{eqnarray}
-f_{2} &=&ig/\left( 2\kappa \right) f_{1}, \\
-gf_{1} &=&\sqrt{2}\kappa \left( -iA+iB\right) , \\
-\sqrt{2}f_{2} &=&A+B, \\
-\sqrt{2}f_{N+1} &=&i^{N-1}A+\left( -i\right) ^{N-1}B, \\
-gf_{N+2} &=&\sqrt{2}\kappa \lbrack i^{N}A+\left( -i\right) ^{N}B],
\end{eqnarray}%
where we obtain $A=0$, $B=-\sqrt{2}f_{2}$, and%
\begin{eqnarray}
f_{1} &=&(2i\kappa /g)f_{2}, \\
f_{l} &=&-\sqrt{2}\left( -i\right) ^{l}f_{2},(3\leqslant l\leqslant N), \\
f_{N+1} &=&\left( -i\right) ^{N-1}f_{2}, \\
f_{N+2} &=&\left( -i\right) ^{N}(2\kappa /g)f_{2}.
\end{eqnarray}%
Note that $A=0$ indicates that $\left\vert \varphi _{\pi /2}\right\rangle $
is a unidirectional plane wave with left-going part $e^{-ikj}$ only.

Similarly, the wave function $\left\vert \tilde{\varphi}_{\pi
/2}\right\rangle $ of Hamiltonian $H^{\dagger }$ for $k=\pi /2$ at $\gamma
=g^{2}/(2\kappa )$ under the Schr\"{o}dinger equations yields
\begin{eqnarray}
-\tilde{f}_{2} &=&-ig/(2\kappa )\tilde{f}_{1}, \\
-g\tilde{f}_{1} &=&\sqrt{2}\kappa (-i\tilde{A}+i\tilde{B}), \\
-\sqrt{2}\tilde{f}_{2} &=&\tilde{A}+\tilde{B}, \\
-\sqrt{2}\tilde{f}_{N+1} &=&i^{N-1}\tilde{A}+\left( -i\right) ^{N-1}\tilde{B}%
, \\
-g\tilde{f}_{N+2} &=&\sqrt{2}\kappa \lbrack i^{N}\tilde{A}+\left( -i\right)
^{N}\tilde{B}],
\end{eqnarray}%
where we obtain $\tilde{A}=-\sqrt{2}\tilde{f}_{2}$, $\tilde{B}=0$, and%
\begin{eqnarray}
\tilde{f}_{1} &=&-(i2\kappa /g)\tilde{f}_{2}, \\
\tilde{f}_{l} &=&-\sqrt{2}i^{l}\tilde{f}_{2},(3\leqslant l\leqslant N), \\
\tilde{f}_{N+1} &=&i^{N-1}\tilde{f}_{2}, \\
\tilde{f}_{N+2} &=&i^{N}(2\kappa /g)\tilde{f}_{2},
\end{eqnarray}%
therefore, the overlap is $\left\langle \tilde{\varphi}_{\pi /2}\right\vert
\varphi _{\pi /2}\rangle =\sum_{l}\tilde{f}_{l}^{\ast }f_{l}$, i.e.,
\begin{equation}
\left\langle \tilde{\varphi}_{\pi /2}\right\vert \varphi _{\pi /2}\rangle =-%
\frac{4\kappa ^{2}}{g^{2}}+1+2\sum_{l=3}^{N}\left( -1\right) ^{l}-\left(
-1\right) ^{N}+\left( -1\right) ^{N}\frac{4\kappa ^{2}}{g^{2}},
\end{equation}%
for even $N$, $\left\langle \tilde{\varphi}_{\pi /2}\right\vert \varphi
_{\pi /2}\rangle =0$; for odd $N$, $\left\langle \tilde{\varphi}_{\pi
/2}\right\vert \varphi _{\pi /2}\rangle =-8\kappa ^{2}/g^{2}\neq 0$. These
indicate that $\gamma =g^{2}/(2\kappa )$ is the exceptional point of the $%
\mathcal{PT}$-symmetric system with even $N$, the coalescence states are the
zero energy states; however, $\gamma =g^{2}/(2\kappa )$ is not the
exceptional point of the $\mathcal{PT}$-symmetric system with odd $N$.

The time evolution of a Gaussian wave packet with central momentum $\pi/2$
is relevant to the near-zero-energy eigen states, i.e., $k\approx \pi /2$,
in which Eq. (\ref{CriticalEQ}) under similar coupling condition ($\kappa
\sim g$) is approximately reduced into
\begin{equation}
\tan \left( Nk\right) \approx 0,
\end{equation}%
the approximate solution of momentum $k$ near $\pi/2$ is%
\begin{equation}
k\approx n\pi /N.
\end{equation}

\section{Dynamics in the $\mathcal{PT}$-symmetric system}

\label{GWPdynamics} We apply the solution of $\mathcal{PT}$-symmetric system
with odd $N$ to calculate the dynamics of a Gaussian wave packet. The gain $%
\gamma =g^{2}/(2\kappa )$ is not the exceptional point of $\mathcal{PT}$%
-symmetric system with odd $N$, thus the biorthoganal basis exists for $H$
and $H^{\dagger }$. We solve the eigen states of $H$ and $H^{\dagger }$ to
form the biorthogonal basis. We use the biorthoganal basis to expand the
initial state and calculate the time evolution. From Eqs. (\ref{EQ1}-\ref%
{EQ3}), we obtain $f_{1}$, $A$, $B$ as functions of $f_{2}$%
\begin{eqnarray}
f_{1} &=&\frac{g}{2\kappa \cos k-i\gamma }f_{2}, \\
A &=&\frac{2i\kappa ^{2}\sin \left( 2k\right) +2\kappa \gamma \sin k-g^{2}}{%
2i\kappa \sqrt{2}e^{2ik}\left( 2\kappa \cos k-i\gamma \right) \sin k}f_{2},
\\
B &=&\frac{2i\kappa ^{2}\sin \left( 2k\right) +2\kappa \gamma \sin k+g^{2}}{%
2i\kappa \sqrt{2}e^{-2ik}\left( 2\kappa \cos k-i\gamma \right) \sin k}f_{2},
\end{eqnarray}%
the coefficients $f_{N+1}$, $f_{N+2}$ can be obtained as functions of $f_{2}$
after substituting $A$, $B$ in Eqs. (\ref{EQ4}-\ref{EQ6}), $H^{\dagger }$ is
solved under the same procedure. The eigen states for $H^{\dagger }$ are set
as $\left\vert \tilde{\varphi}_{k}\right\rangle =\sum_{l}\tilde{f}%
_{l}\left\vert l\right\rangle $ with $\tilde{f}_{l}=\tilde{A}e^{ikl}+\tilde{B%
}e^{-ikl}$ ($3\leqslant l\leqslant N$), similarly, we obtain%
\begin{eqnarray}
\tilde{f}_{1} &=&\frac{g}{2\kappa \cos k+i\gamma }\tilde{f}_{2}, \\
\tilde{A} &=&\frac{2i\kappa ^{2}\sin \left( 2k\right) -2\kappa \gamma \sin
k-g^{2}}{2i\kappa \sqrt{2}e^{2ik}\left( 2\kappa \cos k+i\gamma \right) \sin k%
}\tilde{f}_{2}, \\
\tilde{B} &=&\frac{2i\kappa ^{2}\sin \left( 2k\right) -2\kappa \gamma \sin
k+g^{2}}{2i\kappa \sqrt{2}e^{-2ik}\left( 2\kappa \cos k+i\gamma \right) \sin
k}\tilde{f}_{2},
\end{eqnarray}%
the coefficients $\tilde{f}_{N+1}$, $\tilde{f}_{N+2}$ can be obtained as
functions of $\tilde{f}_{2}$. All coefficients $A$, $B$, $\tilde{A}$, $%
\tilde{B}$\ and wave functions $f_{1}$, $f_{N+1}$, $f_{N+2}$, $\tilde{f}_{1}$%
, $\tilde{f}_{N+1}$, $\tilde{f}_{N+2}$ are expressed as functions of nonzero
values $f_{2}$, $\tilde{f}_{2}$. The momentum $k$ can be solved from the
critical equation\ (\ref{CriticalEQ}) exactly, however, the eigen states
evolved in the time evolution process and being important are the eigen
states in the linear region of the spectrum near $k=\pi /2$, which has
approximate solution of $k\approx n\pi /N$.

Note that the wave functions $\left\vert \varphi _{k}\right\rangle $ and $%
\left\vert \tilde{\varphi}_{k}\right\rangle $ should be renormalized. In a
non-Hermitian system, we use the biorthogonal renormalization method. The
renormalization factors are $G_{k}=\langle \tilde{\varphi}_{k}\left\vert
\varphi _{k}\right\rangle =\sum_{l}\tilde{f}_{j}^{\ast }f_{j}$. The
renormalized eigen states of Hamiltonians $H$ and $H^{\dagger }$ are denoted
as
\begin{eqnarray}
\left\vert \varphi _{G_{k}}\right\rangle
&=&G_{k}^{-1/2}\sum_{l}f_{l}\left\vert l\right\rangle ,  \label{Phik} \\
\left\vert \tilde{\varphi}_{G_{k}}\right\rangle &=&G_{k}^{\ast -1/2}\sum_{l}%
\tilde{f}_{l}\left\vert l\right\rangle ,  \label{Phikprime}
\end{eqnarray}%
for eigenvalues $E_{k}$ and $E_{k}^{\ast }$, respectively. The overlap is $%
\langle \tilde{\varphi}_{G_{k}}\left\vert \varphi _{G_{k}}\right\rangle =1$
according to the definition of $G_{k}$.

The initial state is a Gaussian wave packet with central momentum $\pi /2$,
centered at site $N_{\mathrm{c}}$,
\begin{equation}
\left\vert \psi \left( 0\right) \right\rangle =\Omega
^{-1/2}\sum_{j}e^{-\alpha ^{2}\left( j-N_{\mathrm{c}}\right) ^{2}/2}e^{i(\pi
/2)j}\left\vert j\right\rangle ,
\end{equation}%
the time evolution of the initial state is in form of $\left\vert \psi
\left( t\right) \right\rangle =e^{-iHt}\left\vert \psi \left( 0\right)
\right\rangle $, which can be calculated by employing the biorthogonal basis
$\{\left\vert \varphi _{G_{k}}\right\rangle ,\left\vert \tilde{\varphi}%
_{G_{k}}\right\rangle \}$ that composed by the eigen states of Hamiltonians $%
H$ and $H^{\dagger }$ [Eqs.~(\ref{Phik},~\ref{Phikprime})]. $%
\left\vert \psi \left( t\right) \right\rangle $ in the momentum space of $H$
is in form of
\begin{equation}
\left\vert \psi \left( t\right) \right\rangle =\underset{k}{\sum }%
e^{-iE_{k}t}\langle \tilde{\varphi}_{G_{k}}\left\vert \psi (0)\right\rangle
\left\vert \varphi _{G_{k}}\right\rangle .
\end{equation}%
The Gaussian wave packet is localized in both real and momentum spaces, a
wide Gaussian wave packet in the real space is narrow in the momentum space.
For the Gaussian wave packet with central momentum $\pi /2$, the relevant
eigen states that being important in the time evolution process are the
states with momentum $k$ near $\pi /2$. Therefore, to calculate the
dynamical factor $e^{-iE_{k}t}$, we can approximately express the eigen
energy near momentum $k_{\mathrm{c}}=\pi /2$,
\begin{equation}
E_{k}=-2\kappa \cos k\approx 2\kappa \left( k-k_{\mathrm{c}}\right) .
\end{equation}%
Substituting the expansion coefficients $\langle \tilde{\varphi}%
_{G_{k}}\left\vert \psi (0)\right\rangle $, the time evolution state $%
\left\vert \psi \left( t\right) \right\rangle $ is in form of
\begin{widetext}
\begin{eqnarray}
\left\vert \psi \left( t\right) \right\rangle  &=&\frac{1}{\sqrt{\Omega }}%
\sum_{k}\frac{1}{\sqrt{G_{k}}}(\sum_{j}\tilde{f}_{j}^{\ast }e^{-\frac{\alpha
^{2}\left( j-N_{\mathrm{c}}\right) ^{2}}{2}}e^{ik_{\mathrm{c}%
}j})e^{-i2\kappa t\left( k-k_{\mathrm{c}}\right) }\left\vert \varphi
_{G_{k}}\right\rangle  \\
&\approx &\frac{1}{\sqrt{\Omega }}\underset{k}{\sum }\frac{\tilde{A}^{\ast }%
}{\sqrt{G_{k}}}(\sum_{j}e^{-ikj}e^{-\frac{\alpha ^{2}\left( j-N_{\mathrm{c}%
}\right) ^{2}}{2}}e^{ik_{\mathrm{c}}j})e^{-i2\kappa t\left( k-k_{\mathrm{c}%
}\right) }\left\vert \varphi _{G_{k}}\right\rangle  \\
&=&\frac{\sqrt{2\pi }}{\alpha \sqrt{\Omega }}\underset{k}{\sum }\frac{\tilde{%
A}^{\ast }}{\sqrt{G_{k}}}e^{-i\left( k-k_{\mathrm{c}}\right) N_{\mathrm{c}%
}}e^{-\frac{\left( k-k_{\mathrm{c}}\right) ^{2}}{2\alpha ^{2}}}e^{-i2\kappa
t\left( k-k_{\mathrm{c}}\right) }\left\vert \varphi _{G_{k}}\right\rangle
\end{eqnarray}%
then we substitute the eigen states $\left\vert \varphi
_{G_{k}}\right\rangle =\sum_{j}(f_{j}/\sqrt{G_{k}})\left\vert j\right\rangle
$ in the expression of $\left\vert \psi \left( t\right) \right\rangle $, in
the real space, we get
\begin{eqnarray}
\left\vert \psi \left( t\right) \right\rangle  &\approx &\frac{\sqrt{2\pi }}{%
\alpha \sqrt{\Omega }}\sum_{j=3}^{N}\sum_{k}\frac{\tilde{A}^{\ast }}{G_{k}}%
e^{-i\left( k-k_{\mathrm{c}}\right) N_{\mathrm{c}}}e^{-\frac{\left( k-k_{%
\mathrm{c}}\right) ^{2}}{2\alpha ^{2}}}e^{-i2\kappa t\left( k-k_{\mathrm{c}%
}\right) }f_{j}\left\vert j\right\rangle  \\
&=&\sqrt{\frac{2\sqrt{\pi }}{\alpha }}\sum_{j=3}^{N}\sum_{k}\frac{\tilde{A}%
^{\ast }B}{G_{k}}e^{-ikj}e^{-i\left( k-k_{\mathrm{c}}\right) N_{\mathrm{c}%
}}e^{-\frac{\left( k-k_{\mathrm{c}}\right) ^{2}}{2\alpha ^{2}}}e^{-i2\kappa
t\left( k-k_{\mathrm{c}}\right) }\left\vert j\right\rangle
\label{WaveEmission} \\
&&+\sqrt{\frac{2\sqrt{\pi }}{\alpha }}\sum_{j=3}^{N}\sum_{k}\frac{\tilde{A}%
^{\ast }A}{G_{k}}e^{ikj}e^{-i\left( k-k_{\mathrm{c}}\right) N_{\mathrm{c}%
}}e^{-\frac{\left( k-k_{\mathrm{c}}\right) ^{2}}{2\alpha ^{2}}}e^{-i2\kappa
t\left( k-k_{\mathrm{c}}\right) }\left\vert j\right\rangle ,
\label{ReflectedWP}
\end{eqnarray}%
for the center part of the system (from $j=3$ to $j=N$). The time evolution
of the Gaussian wave packet $\left\vert \psi \left( t\right) \right\rangle $
is reduced to a combination of two parts. After simplification, we note that
before the Gaussian wave packet reaching the gain site of the $\mathcal{PT}$%
-symmetric system, Eq. (\ref{ReflectedWP}) dominates and approximately
equals to the time evolution of the Gaussian wave packet on a uniform chain;
after the Gaussian wave packet being reflected at the gain site [$%
t>t_{0}\approx (N-N_{\mathrm{c}})/(2\kappa )]$, Eq. (\ref{WaveEmission})
represents a persistent wave emission and Eq. (\ref{ReflectedWP}) stands for
a reflected wave packet. The platform height $h=2(\gamma _{\mathrm{c}%
}/\kappa )^{2}\sqrt{\pi }/\alpha $ shown in Eq. (\ref{h}) is obtained after
simplifying Eq. (\ref{WaveEmission}), this persistent wave
emission part results in a linear increasing of the initial state probability. As long as the system is at critical $\gamma _{\mathrm{c}}$,
the linear increasing of the probability is visible even if the gain is
small. Moreover, when $g^{4}/\kappa ^{4}\gg 2\alpha ^{2}/\pi $, the wave
emission dominates in the time evolution process. $\left\vert \psi \left(
t\right) \right\rangle $ is approximately described by a platform wave emission , as
\begin{eqnarray}
\left\vert \psi \left( t\right) \right\rangle  &\approx &\sqrt{\frac{2\sqrt{%
\pi }}{\alpha }}\sum_{j=3}^{N}\sum_{k}\frac{\tilde{A}^{\ast }B}{G_{k}}%
e^{-ikj}e^{-i\left( k-k_{\mathrm{c}}\right) N_{\mathrm{c}}}e^{-\frac{\left(
k-k_{\mathrm{c}}\right) ^{2}}{2\alpha ^{2}}}e^{-i2\kappa t\left( k-k_{%
\mathrm{c}}\right) }\left\vert j\right\rangle  \\
&\approx &\frac{1}{4}\left( \frac{g}{\kappa }\right) ^{2}\sqrt{\frac{2\sqrt{%
\pi }}{\alpha }}\sum_{j=3}^{N}\left( \underset{k}{\sum }\frac{i\left(
N-2\right) ^{-1}}{\left( k-k_{\mathrm{c}}\right) }e^{4ik}e^{-i\left( k-k_{%
\mathrm{c}}\right) \left( N_{\mathrm{c}}+2\kappa t\right) }e^{-\frac{\left(
k-k_{\mathrm{c}}\right) ^{2}}{2\alpha ^{2}}}e^{-ikj}-e^{-ik_{\mathrm{c}%
}j}\right) \left\vert j\right\rangle  \\
&=&\frac{1}{2}\sqrt{\frac{2\sqrt{\pi }\gamma _{_{\mathrm{c}}}^{2}}{\alpha
\kappa ^{2}}}\sum_{j=3}^{N}\left( \underset{k}{\sum }i\frac{\left(
N-2\right) ^{-1}}{\left( k-k_{\mathrm{c}}\right) }e^{-\frac{\left( k-k_{%
\mathrm{c}}\right) ^{2}}{2\alpha ^{2}}}e^{-i\left( k-k_{\mathrm{c}}\right)
\left( j+N_{\mathrm{c}}+2\kappa t-4\right) }-1\right) e^{-ik_{\mathrm{c}%
}j}\left\vert j\right\rangle  \\
&=&\frac{\sqrt{h}}{2}\sum_{j=3}^{N}\left( \sum_{k}i\frac{\left( N-2\right)
^{-1}}{\left( k-k_{\mathrm{c}}\right) }e^{-\frac{\left( k-k_{\mathrm{c}%
}\right) ^{2}}{2\alpha ^{2}}-i\left( k-k_{\mathrm{c}}\right) \left( j+N_{%
\mathrm{c}}+2\kappa t-4\right) }-1\right) e^{-ik_{\mathrm{c}}j}\left\vert
j\right\rangle  \\
&\approx &-\frac{\sqrt{h}}{2}\sum_{j=3}^{N}\left\{ \text{\textrm{erf}}\left(
\frac{2^{3/4}}{\Delta }\left[ 2(\kappa t-N-2)+j+N_{\mathrm{c}}\right]
\right) +1\right\} e^{-\frac{i\pi j}{2}}\left\vert j\right\rangle , \\
&=&-\frac{\sqrt{h}}{2}\sum_{j=3}^{N}\left\{ \text{\textrm{erf}}\left[ \frac{%
2^{3/4}}{\Delta }(j-N_{t})\right] +1\right\} e^{-\frac{i\pi j}{2}}\left\vert
j\right\rangle .
\end{eqnarray}%
\end{widetext}
Where $N_{t}=2(N+2)-N_{\mathrm{c}}-2\kappa t$ indicates the Gaussian wave
packet center after reflection for system with $\gamma=0$. In the summation of $k$, $%
k\approx n\pi /N$ with integer $n=1,2,\cdots ,N$. The time evolution state $%
\left\vert \psi \left( t\right) \right\rangle $ at the end with gain is $%
\left\vert \psi \left( N+1,t\right) \right\rangle =-(\sqrt{2}i\gamma \kappa
/g^{2})\left\vert \psi \left( N,t\right) \right\rangle $, and $\left\vert
\psi \left( N+2,t\right) \right\rangle =-(\sqrt{2}\kappa /g)\left\vert \psi
\left( N,t\right) \right\rangle $. In the $\mathcal{PT}$-symmetric system
described by Hamiltonian in Eq.~(\ref{H_PT}), the total site number of the
system is $N+2$. However, the time evolution discussed in Sec. \ref{Emission}
is for a system with gain only at one end, the total site number is $N+1$.
In that case, therefore, we obtain the time evolution of the Gaussian wave
packet in the wave emission system as shown in Eq.~(\ref{Phi_t}).

\end{document}